\documentclass[12pt]{article}

\usepackage{full page}			% to increase margins
\usepackage{graphicx}			% to add graphics
\usepackage{authblk}			% to add author's affiliation
\usepackage{natbib}				% for bibliography
\usepackage{amsmath, amssymb, amsfonts, amsthm, bbm, bm, mathrsfs}  % for mathematical symbols
\usepackage{color}              % to customize color
\usepackage{graphicx}           % to add graphics
\usepackage{enumerate}			% to customize lists
\usepackage{booktabs}			% to customize tables
\usepackage{tikz}   			% to draw figures
\usepackage{here}				% to force the picture location
\usepackage{url, hyperref}		% to write urls and add hyperlinks

\usetikzlibrary{arrows,decorations,snakes} % tikz libraries used

\usepackage{caption, subcaption} % to have figures and subfigures
\usepackage[labelformat=simple]{subcaption}
\graphicspath{{../figures/}}

\begin{document}

\title{Stop or Continue Data Collection: A Nonignorable Missing Data Approach for Continuous Variables}
%\author{}
\author[1]{Thais Paiva}
\author[2]{Jerome P. Reiter}
\affil[1]{\small Department of Statistics, Federal University of Minas Gerais, Belo Horizonte, Brazil}
\affil[2]{Department of Statistical Science, Duke University, Durham, NC, USA}

%\date{\today}
\maketitle

\begin{abstract}
We present an approach to inform decisions about nonresponse follow-up sampling. The basic idea is (i) to create completed samples by imputing nonrespondents' data under various assumptions about the nonresponse mechanisms, (ii) take hypothetical samples  of varying sizes from the completed samples, and (iii) compute and compare measures of accuracy and cost for different proposed sample sizes. As part of the methodology, we present a new approach for generating imputations for multivariate continuous data with nonignorable unit nonresponse. We fit mixtures of multivariate normal distributions to the respondents' data, and adjust the probabilities of the mixture components to generate nonrespondents' distributions with desired features. We illustrate the approaches using data from the 2007 U. S. Census of Manufactures.
\\

\end{abstract}

%------------------------------------------------------------------------------

\section{Introduction} \label{sec:introduction}

With shrinking budgets and expanding costs, government agencies and survey organizations must find ways to reduce the expenses of data collection. One promising set of approaches is adaptive design, in which the agency changes the design of the survey based on the characteristics of the respondents and any known information about the nonrespondents. For example, the agency may focus follow-up efforts on individuals with characteristics not adequately represented among the respondents \citep[e.g.,][]{Wag08, Sch09, Sch11, Sch12}, undertake a systematic subsample of nonrespondents \citep[e.g.,][]{Kap14}, or cease follow-up efforts for individuals deemed highly unlikely to respond without excessive expense \citep[e.g.,][]{Mil13, Fin13}. 

In this article, we consider one potential decision point in adaptive designs: should the agency stop all efforts at data collection, given what has been already obtained? Stopping data collection reduces costs and allows for earlier release of data, but doing so only makes sense when data quality is not sacrificed to undesirable levels. To make informed decisions, agencies can benefit from principled approaches for assessing the trade offs in projected costs and estimated accuracy at specific times in the data collection process.

Several methods for stopping data collection have been suggested in the literature. Most relevant to our work, \citet{Rao08} and \citet{Wag10} propose stopping rules for surveys with multiple waves of nonresponse follow-up and a binary response variable. Their basic framework is to impute the missing values for all individuals who have yet to respond by wave $t$, evaluate some estimand on the completed data, and stop when the change in the estimate from wave $t-1$ to wave $t$ is small according to some criterion. \citet{Rao08} and \citet{Wag10} use imputation models that presume the nonrespondents at wave $t$ follow the same distribution as respondents up to (or possibly only in) wave $t-1$; that is, the nonrespondents' data are missing at random (MAR) \citep{Rub76}. While convenient, MAR is a strong assumption that may be unrealistic, resulting in unreliable imputations and, consequently, ineffective stopping decisions. 

In this article, we propose an imputation-based approach to define stopping rules when the key variables of interest are continuous in nature. Unlike \citet{Rao08} and \citet{Wag10}, we create completed datasets without assuming MAR; that is, we develop and use a multivariate model for nonignorable unit nonresponse. To do so, we fit mixtures of multivariate normal distributions to the respondents' data, and adjust the probabilities of the mixture components (keeping fixed the location and scale parameters) to generate nonrespondents' distributions with desired features. For example, to represent a scenario in which nonrespondents are more (less) likely than respondents to come from a particular region of support, we inflate (deflate) the probabilities of the components with high density in that region. These adjustments facilitate sensitivity analysis---arguably the best one can do with nonignorable missing data---as the agency can evaluate inferences on completed data constructed with different sets of altered probabilities. With such evaluations, agencies can decide whether or not the potential cost savings are worth the loss in accuracy from stopping.  

The remainder of this article is organized as follows. In Section \ref{sec:methodology}, we describe how we adapt mixtures of multivariate normal distributions to handle nonignorable unit nonresponse. We note that this approach could be used for general imputation tasks with multivariate continuous data; it is not restricted to stopping rule decisions. In Section \ref{sec:imp_census}, we illustrate how to apply this methodology on a subset of data from the 2007 U. S. Census of Manufactures. In Section \ref{sec:met_adesign}, we describe how to use the nonignorable missing data procedure and sensitivity analysis to inform stopping decisions. We define several accuracy metrics that can aid the decision process. In Section \ref{sec:application}, we illustrate the stopping rule methodology on the data  from the 2007 U. S. Census of Manufactures. We demonstrate the methodology using one wave of nonresponse follow-up. Finally, in Section \ref{sec:conclusions} we conclude with a discussion and suggestions for future research.

%------------------------------------------------------------------------------

\section{Model for Multivariate Continuous Data with Nonignorable Nonresponse} \label{sec:methodology}

To fix notation, consider an agency that intends to collect data on $n$ individuals, either in a census or a sample. For $i= 1, \dots, n$, let $y_{ij}$ denote the value of variable $j$ for individual $i$. Let $\bm{y}_i = (y_{i1}, \dots, y_{ip})$ include the values of the $p$ variables of interest. Let $\bm Y = ({\bm{y}_1, \dots, \bm{y}_{n}})'$.  At any time $t$ in the data collection period, let $r_{it} = 1$ if individual $i$ has provided data by time $t$, and let $r_{it}=0$ if individual $i$ has not yet responded. Let $\bm{R}_t = (r_{1t}, \dots, r_{nt})$. We only consider unit nonresponse when defining each $r_{it}$. Let $\bm{Y}_t$ comprise the data for the $n_t$ individuals with $r_{it}=1$. We note that $(\bm{R}_t, \bm{Y}_t)$ is updated as $t$ advances. At any time $t$, our goal is to impute values of $\bm Y$ for the nonrespondents, allowing for possibly nonignorable missingness. 

When handling multivariate data that are not missing at random (NMAR), analysts generally take one of two approaches. In selection models, analysts posit a joint model for $\bm Y$ and a model for $(\bm{R}_t \mid \bm Y)$, such as a logistic regression \citep{Gly86, Rub87, Dig94, Mol97}. In pattern mixture models, analysts posit a model for $\bm{R}_t$ and a model for $(\bm Y \mid \bm{R}_t)$; that is, there are different models for $\bm Y$ for respondents and nonrespondents \citep{Gly93, Lit93, Lit94, Mol98b, Thi02}. We adopt the pattern mixture approach, as we find it easier to utilize for sensitivity analysis; see \citet{Lit02}, \citet{Lit95,Lit08}, \citet{Dan08}, \citet{And11}, and \citet{And15} for further discussion.

Pattern mixture models can be implemented in two steps. First, the analyst fits a model to the observed data, say $f(\bm{Y}_t)$. Second, the analyst imposes departures from $f(\bm{Y}_t)$, for example by altering certain model parameters, that reflect beliefs about possible distributions for the nonrespondents. The analyst can generate multiple imputations \citep{Rub87} for nonrespondents to create completed versions of $\bm{Y}$ for inference. By examining multiple possible distributions for the nonrespondents, the analyst can perform sensitivity analysis.

When building off $f(\bm{Y}_t)$, it is beneficial to use a model that accurately describes the multivariate distribution among the respondents. Often this multivariate distribution has complicated features, especially for economic data. We use mixtures of multivariate normal distributions as the engine for flexible modeling of the joint distribution of $\bm{Y}_t$. These models can capture complex relationships among variables, such as non-linearities and interactions, as well as multiple modes, heteroscedasticity, and skewness \citep{Fer83,Gho01,Gho07,Mul13}. These models also have parameters that can be easily manipulated to generate possible distributions for nonrespondents.

\subsection{Mixtures of multivariate normal distributions} \label{sec:met_model}

To begin, we review the mixture normal distribution that we use to model $\bm{Y}_t$ at any time $t$.  To facilitate model specification and estimation, we first standardize each variable in $\bm{Y}$. We assume that each individual in the dataset belongs to one of $K < \infty$ mixture components, indicated by $z_i \in \{1,\dots,K\}$. Let $\bm Z = (z_1, \dots, z_n)$. Let $\pi_k= P(z_i=k)$, where $i=1, \dots, n$ and $k=1, \dots, K$, be the probability that individual $i$ belongs to component $k$. Let $\bm \pi = (\pi_1, \dots, \pi_K)$. Within each component, let $\bm y_i$ follow a multivariate normal distribution with mean $\bm \mu_k$ and variance $\bm \Sigma_k$. Let $\bm \mu = (\bm \mu_1, \dots, \bm \mu_k)$, and let $\bm \Sigma = (\bm \Sigma_1, \dots, \bm \Sigma_k)$. We can write the model as 
\begin{eqnarray}
 \bm y_i | z_i, \bm \mu, \bm \Sigma &\sim& N(\bm \mu_{z_i}, \bm \Sigma_{z_i})
\label{eq:model_yi} \\
 z_i | \bm \pi &\sim& \mbox{Multinomial}(\pi_1,\dots,\pi_K).
\label{eq:model_zi}
\end{eqnarray}
Because of their ability to capture complex distributions, mixture models have been used in applications across a variety of fields \citep[e.g., ][]{Wes93, Esc95, Mul04, Dun10, She13}.

We use a Bayesian approach to estimate $(\bm \mu, \bm \Sigma, \bm \pi)$.  As the prior distribution on $\bm \pi$, we use the stick-breaking representation of a truncated Dirichlet process \citep{Fer73, Fer83, Esc95, Wes94, Set94, Ish01}, which has been shown to be convenient for density estimation \citep{Mul13}. We have  
\begin{eqnarray}
 \pi_k &=& v_k \, \textstyle \prod_{g<k} (1-v_g) \quad \mbox{for }
k=1,\dots,K \label{eq:prior_pi} \\
 v_k &\sim& \mbox{Beta}(1,\alpha) \quad \mbox{for } k=1,\dots,K-1; \; v_K = 1
\label{eq:prior_v} \\
 \alpha &\sim& \mbox{Gamma}(a_{\alpha}, b_{\alpha}). \label{eq:prior_alphaDP}
\end{eqnarray}
Following \citet{Kim14}, we set $a_\alpha = b_\alpha = 0.25$ to represent a small effective sample size in the Gamma distribution for $\alpha$. This allows the likelihood function from the data to dominate the prior distribution for $\bm \pi$. For further discussion about this prior distribution, see \citet{Dun09} and \citet{Kim14}. 

For the prior specification of $(\bm \mu, \bm \Sigma)$, we consider two alternatives. The most general uses  
\begin{eqnarray}
 \bm \mu_k | \bm \Sigma_k &\sim& N(\bm \mu_0, h^{-1} \bm \Sigma_k) \label{eq:prior_muDP}\\
 \bm \Sigma_k &\sim& \mbox{InverseWishart}(f, \Phi) \label{eq:prior_sigmaDP}.
\end{eqnarray}
Here, $f$ is a prior degrees of freedom and $\Phi = diag(\phi_1,\dots,\phi_p)$, with $\phi_j \sim \mbox{Gamma}(a_{\phi},b_{\phi})$ for $j=1,\dots,p$. We specify priors with $a_{\phi} = b_{\phi} = 0.25$ to allow substantial prior mass at modest sized variances. We set $\bm \mu_0=0$, since the variables are standardized, $f=p+1$ to ensure a proper posterior distribution, and $h=1$ for convenience. A second alternative, and the one we use in the Census of Manufactures application, replaces \eqref{eq:prior_sigmaDP} with fixed spherical covariance matrices, $\bm \Sigma_k = \sigma \bm I_p$ for all $k$ and some agency-specified value of $\sigma>0$. In general, using equal $\bm \Sigma_k$ with a small $\sigma$ requires more components to describe the distribution adequately than allowing arbitrary $\bm \Sigma_k$. The number of required components increases as $\sigma$ gets small.

With either specification for $(\bm \mu, \bm \Sigma)$, we can use a standard
Gibbs sampler to estimate the posterior distribution; see the supplementary material for details. The Gibbs sampler algorithms are easily modified to handle missing values in $\bm y_i$ that are MAR. Given a current draw of $(\bm \mu, \bm \Sigma, \bm Z)$, we take a draw of the missing values in $\bm y_i$ from the conditional normal distribution given the observed values in $\bm y_i$. We use $N(\bm \mu_{z_i}, \bm \Sigma_{z_i})$ to derive the relevant conditional distribution.   

To specify $K$, we follow the recommendations from \citet{Kim15}. We recommend starting with a somewhat large value, for example $K = 30$. In each iteration of the Gibbs sampler, we recommend counting the number of components that include at least one $\bm y_i$. If this count ever reaches $K$, it is prudent to increase $K$ and refit the model with more components. When the count is always less than $K$, i.e., in every iteration some components are empty, then the value of $K$ is reasonable.  One also could identify reasonable values of $K$ by fitting the mixture models on similar data collected in prior years or previous waves of a longitudinal study.

In many datasets, the data values are non-negative and sometimes equal to zero.  As described by \citet{Kim15}, one can fit the mixture model to the logarithms, which ensures positive imputations, adding a small constant to zero values before taking the logarithms. Alternatively, one could use a mixture model for mixed data types as in \cite{Mur16}.

\subsection{Pattern mixture modeling} \label{sec:met_imputation}

When $\bm Y$ suffers from unit nonresponse, we can use the model in \eqref{eq:model_yi}--\eqref{eq:prior_sigmaDP}, possibly with $\bm \Sigma_k = \sigma \bm I_p$ for all $k$, to impute values for the $n_0 = n - n_t$ cases with $r_{it}=0$. At every iteration in the Gibbs sampler, we draw a value of $z_i$ from \eqref{eq:model_zi} for each case with  $r_{it}=0$, and draw a value of $\bm y_i$ from the corresponding $N(\bm \mu_{z_i}, \bm \Sigma_{z_i})$. This generates imputed values from the same distribution as $\bm Y_{t}$; that is, the unit nonresponse is assumed to be MAR.  

To construct pattern mixture models for NMAR unit nonresponse, we again estimate the mixture normal model with $\bm Y_t$. In each iteration of the Gibbs sampler, we evaluate the posterior density from \eqref{eq:model_yi}--\eqref{eq:prior_sigmaDP}. Let $(\bm \mu^m, \bm \Sigma^m)$ equal the values of $(\bm \mu, \bm \Sigma)$ at the iteration with the maximum posterior value (MAP) of the posterior density \citep{Fra07}; note that $\bm \Sigma^m = \sigma \bm I_p$ when using the equal covariance model. Let $\bm \pi^m$ equal the value of $\bm \pi$ at the MAP iteration. Rather than draw $z_i$ for the $n_0$ nonrespondents using $\bm \pi^m$, which would imply a MAR imputation process, we define and use new mixture probabilities  $\bm \pi^* = (\pi^*_1, \dots, \pi^*_K)$. These are chosen to reflect beliefs about the nonrespondents' distribution, as we illustrate in Section \ref{sec:imp_census}. We then generate the unit nonrespondents' values from  
\begin{eqnarray}\label{eq:imp_model_zi} 
z_i | \bm \pi^*, r_{it}=0 &\sim& \text{Multinomial}(\pi^*_1,\dots,\pi^*_K) \\
{\bm y}_{i} | z_i, \bm \mu^m, \bm \Sigma^m, r_{it}=0 &\sim& N(\bm
\mu_{z_i}^m, \bm \Sigma_{z_i}^m).\label{eq:imp_model_yi}
\end{eqnarray}

By fixing $(\bm \mu^m, \bm \Sigma^m)$ and altering $\bm \pi^m$, we assume the nonrespondents belong to the same components as $\bm Y_t$ but at different frequencies. This allows the agency to upweight or downweight regions of support evident in $\bm Y_t$ when specifying the nonrespondents' distribution. For example, analysts can set $\pi^*_k$ to be small (or zero) for components with most of the density over regions of $\bm Y$ that are deemed improbable for nonrespondents. Or, analysts can increase $\pi^*_k$ for components with density in regions that are deemed more probable than observed in $\bm Y_t$. The closer $\bm \pi^*$ is to $\bm \pi^m$, the closer the data are to MAR.  

Specifying the entire vector of probabilities can be complicated in some situations, particularly with large $p$. To facilitate the process of setting $\bm \pi^*$, we developed the NIMC (Nonignorable missingness Imputation for Multivariate Continuous data) tool. The NIMC is an R \citep{R} application developed with the \verb|shiny| package \citep{Shiny} that allows users to alter components' probabilities via a graphical interface. The NIMC starts at the MAR model, i.e., using $\bm \pi^{m}$, allowing the user to create and examine departures from MAR by increasing or decreasing any $\pi_k^m$. The NIMC renorms the altered probabilities of all components to sum to one, to ensure the resulting mixture model is a proper density. Using the renormed probabilities, the application automatically generates imputations for the missing cases following the model described in \eqref{eq:imp_model_zi} and \eqref{eq:imp_model_yi}, so that the analyst can change $\bm \pi^*$ to arrive at desired distributions. An illustrative version of the NIMC tool is available at 
%({\em URL to website not provided to ensure blinded review} 
\url{http://sites.duke.edu/tcrn/research-projects/downloadable-software/}. 

By controlling $\bm \pi^*$, the analyst controls where the nonrespondents are likely to fall in the space defined by the respondents. We recommend adjusting ten to fifteen values of $\pi_k^m$, and setting the remainder either to zero or to a renormalized version of $\pi_k^m$. In our experience, this provides enough flexibility to generate different types of nonrespondent distributions without becoming so large as to make specification of $\bm \pi^*$ too cumbersome.

\subsection{Implementing this pattern mixture approach} \label{sec:met_implementation}

Before illustrating the approach on the Census of Manufactures data, we discuss some general implementation issues, in particular the choice of unequal or equal $\bm \Sigma_k$, what to do with unoccupied components, steps to facilitating interpretations, and what to do when nonrespondents are unlike any respondents.  

%\subsubsection{Choosing unequal or equal $\Sigma_k$} 
For some $\bm Y_t$, the model with unequal $\bm \Sigma_k$ results in only a small number of occupied components or results in many components with substantially overlapping regions of high density. For such $\bm Y_t$, it can be difficult to alter $\bm \pi^m$ in ways that enable meaningful sensitivity analysis, as it is difficult to generate different distributions for nonrespondents and respondents by adjusting $\bm \pi^m$. In these cases, we recommend setting $\bm \Sigma_k = \sigma \bm I_p$ for all $k$, and trying multiple values of $\sigma$ until finding one that results in fitted components amenable to creating NMAR missing data scenarios. In our experience, when modeling standardized $\bm y_i$ we find that using $ 0< \sigma < 1$ results in usable numbers of occupied components. 

%\subsubsection{What to do with unoccupied components}
When the specified $K$ is large, we expect some components in the MAP iteration to be empty. We suggest setting $\pi_k^* = 0$ for all empty components. The $(\bm \mu_k^m, \bm \Sigma_k^m)$ for empty components is derived entirely from the prior distribution. Such $(\bm \mu_k^m, \bm \Sigma_k^m)$ are not useful for generating nonrespondents' values in a pattern mixture model approach.

%\subsubsection{Facilitating interpretations} 
To facilitate setting $\bm \pi^*$, it can be useful to rank the components after finding $(\bm \mu^m, \bm \Sigma^m, \bm  \pi^m)$. In the Census of Manufactures application, we rank components based on their distance to an artificially created point with small values of all variables. Specifically, we rank on $\Delta_k = (\bm \mu_k - \bm y_{\min})'(\bm \mu_k - \bm y_{\min})$, where $\bm y_{\min}=\left(\min_i(y_{11}, \dots, y_{n1}),\dots, \min_i(y_{1p}, \dots, y_{np})\right)$. Rankings facilitate interpretation of the components. For example, to impute more nonrespondents' data with large values of $\bm y_i$, we inflate the $\pi^m_k$ of the components with the highest ranks on $\Delta_k$. To generate nonrespondents' data with low values of $\bm y_i$ we inflate the $\pi^m_k$ of the components with lowest ranks on $\Delta_k$. 

%\subsubsection{Nonrespondents unlike any respondents.}
Even though changing $\bm \pi^m$ results in different respondents' and nonrespondents' distributions, the nonrespondents' values are imputed in the same regions as $\bm Y_t$, albeit with different propensities. Adjusting $\bm \pi^m$ does not capture NMAR settings where the missing values are suspected to be in regions with no respondents' data. To represent such scenarios, the analyst must create new components. This requires specifying new probabilities and the mean and covariance matrix of the new components.

\subsection{Incorporating auxiliary information}

In some databases, the agency may possess external information on respondents and nonrespondents, such as paradata or administrative records.  In the Census of Manufactures, for example, data on establishments' employment, sales, and payroll are available to the Census Bureau from the U.\ S.\ Internal Revenue Service. When these administrative values are highly correlated with the recorded values, the agency could use the auxiliary data to improve imputations and potentially make more accurate decisions. 

When the number of auxiliary variables is modest, the agency can include these variables in $\mathbf{y}_i$ and estimate the model in \eqref{eq:model_yi} and \eqref{eq:model_zi}. It is straightforward to generate imputations for nonrespondents under MAR. For each nonrespondent $i$, sample its $z_i$ from its conditional distribution given the auxiliary values. Then, given a sample of $z_i$, sample the missing values in $\mathbf{y}_i$ from the appropriate conditional multivariate normal distribution for that $z_i$.  When the number of auxiliary variables is large, it can be preferable to include these variables as regressors in the component means. 

While it may be reasonable to impute missing values for establishments and variables with administrative data under MAR assumptions, agencies still may want to assess sensitivity to nonignorable missingness mechanisms when using auxiliary variables. One way to do so is to modify the algorithm in Section \ref{sec:met_imputation} as follows. We specify $\boldsymbol{\pi}^*$ as before, letting it reflect beliefs about the distributions of nonrespondents. However, rather than sample $z_i$ from \eqref{eq:imp_model_zi}, we sample it from the conditional distribution of $z_i$ given the auxiliary variables for nonrespondent $i$ and $\boldsymbol{\pi}^*$. We sample the missing values in $\mathbf{y}_i$ from the appropriate conditional multivariate normal distribution for that $z_i$.  This allows the auxiliary data to influence the component assignment for record $i$, while adjusting the $\boldsymbol{\pi}_i$ to reflect NMAR assumptions.

We do not consider settings with auxiliary data for the remainder of the paper. Investigating how best to use auxiliary data is a topic worthy of future research.

%------------------------------------------------------------------------------

\section{Imputation of Nonignorable Unit Nonresponse in the Census of Manufactures}\label{sec:imp_census}

We now illustrate the proposed approach to pattern mixture modeling using data from the 2007 U.\ S.\ Census of Manufactures (CMF). The CMF covers all establishments in manufacturing industries in the United States. Data are collected via mail-in questionnaires \citep{ASM_online}. We focus on three highly important variables in the CMF, namely the total value of shipments (TVS), total employment (TE), and salary/wages (SW). We use data from the ready-mix concrete industry, which is relatively homogeneous in these three variables as most of their business is based on a single product. We also examined the plastics products manufacturing industry, which is a more broad classification and more heterogeneous. The results for the plastics products industry are included in the supplementary material.

CMF data are collected and processed in waves during the year, depending on when each establishment mails its form. This can result in NMAR unit nonresponse at some waves. For example, larger establishments may employ individuals who are intimately familiar with and responsible for the types of variables requested in the CMF, so that it is straightforward for them to return the questionnaire. On the other hand, smaller companies may not have such employees, so that it takes them longer to get the right data and return the completed CMF forms. The CMF data include response date, so that we can tell when establishments returned their questionnaires, if at all. 

For illustration, we consider as $\bm Y_t$ all the establishments that had at least one reported value among the three variables of interest and that had a valid response date. We consider the remaining establishments as unit nonrespondents. Some establishments in the file have data on the three key variables but no response date, which could be due to miscellaneous reasons, including late response. For purposes of illustration, we consider these cases as unit nonrespondents and impute their entire vector of survey variables. Of course, in an actual application rather than an illustration, we would use these partially observed data and impute only the missing items. Due to disclosure limitations imposed by the Census Bureau, we cannot provide the sample sizes.

We consider three scenarios when setting $\bm \pi^*$. The first is to assume MAR nonresponse, in which we set $\bm \pi^* = \bm \pi^m$. The second scenario assumes that nonrespondents tend to be small establishments. To generate nonrespondents' data consistent with this scenario, we inflate $\pi^m_k$ for components with small values of $\Delta_k$. The third scenario represents an opposite NMAR mechanism: we inflate $\pi^m_k$ for components with large values of $\Delta_k$. This scenario is in some ways a worst case scenario, since large establishments can have disproportionate impacts on summary statistics (e.g., totals) typically computed with CMF data. For all scenarios, we use the mixture normal model with $\bm \Sigma_k = \sigma_p \bm I_p$ for all $k$, where $\sigma = 0.3$. Based on investigations with different values of $\sigma$, the value 0.3 offers sufficient flexibility to impute data in all regions of the variable space while being large enough to make choosing $\bm \pi^*$ not cumbersome. The MAP iteration has 17 non-empty components.

\begin{figure}[t]
 \centering
 \includegraphics[width=\textwidth]{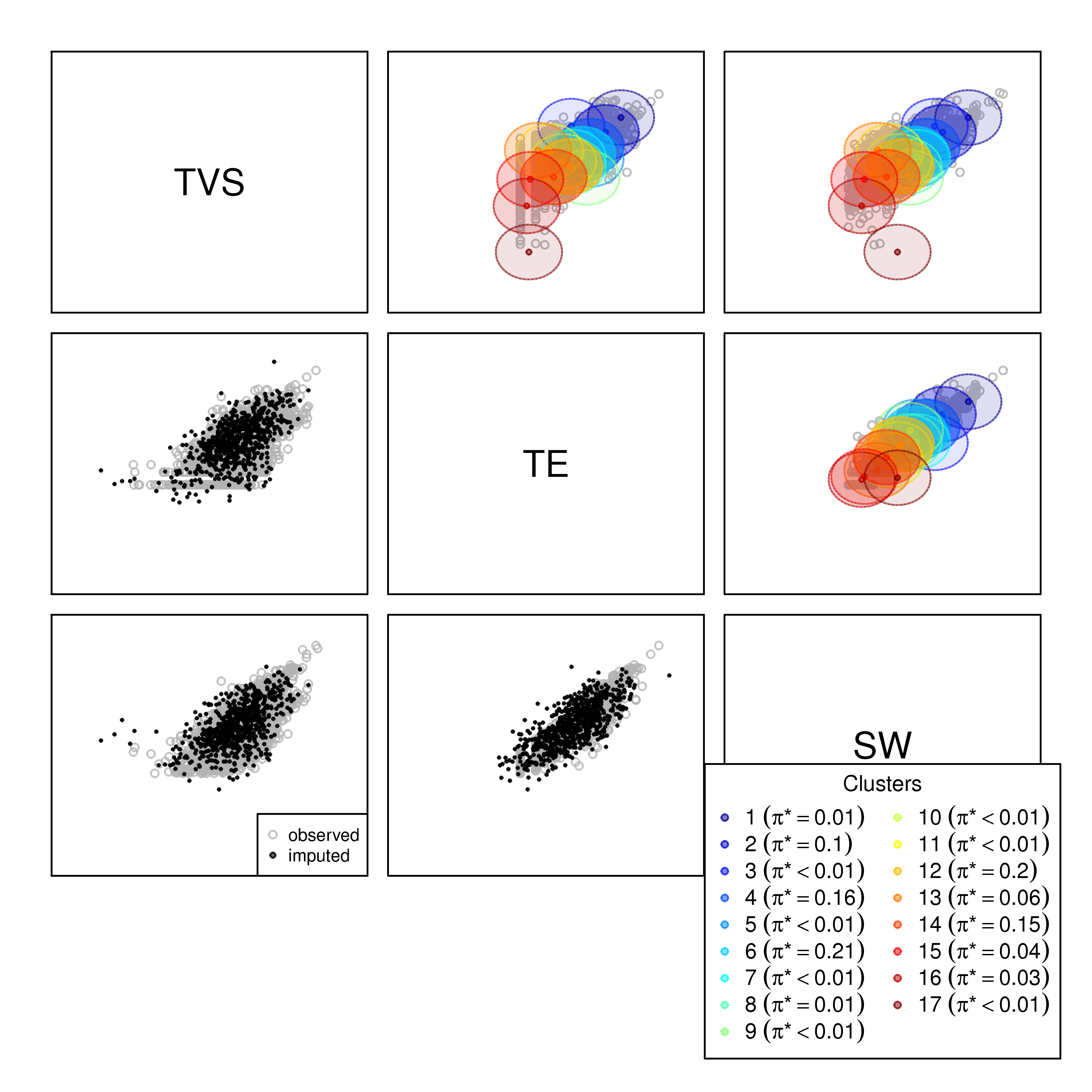}
 \caption[Pairwise scatter plots with the results for the Concrete industry from the 2007 CMF, for the MAR imputation scenario.]{\label{fig:co_MAR} Pairwise scatter plots for the concrete industry from the 2007 CMF for the MAR imputation scenario. Observed points are plotted as gray hollow circles. The black filled circles on the lower diagonal are the imputed points. The colored circles on the upper diagonal are the 95\% quantile ellipses of the fitted clusters, with color intensity proportional to the mixture probabilities. Graph copied from the output of the NIMC R tool. Axes removed to prevent disclosure of information about the magnitude of the data.}
\end{figure}

Because the data are heavily skewed, we log transform and standardize
the variables. We exclude the few cases with zero value for shipments,
employment or salary when estimating the model (alternatively, one could replace the zeros with small positive values). Figure \ref{fig:co_MAR} displays pairwise scatter plots for the concrete industry after imputation using MAR. Most of the density is concentrated in fewer than half of the components. As desired in the MAR scenario, the imputed values for nonresponse have essentially the same distribution as $\bm Y_t$, which suggests that the mixture normal model describes the distribution of $\bm Y_t$ reasonably well.

\begin{figure}[t]
 \centering
 \includegraphics[width=\textwidth]{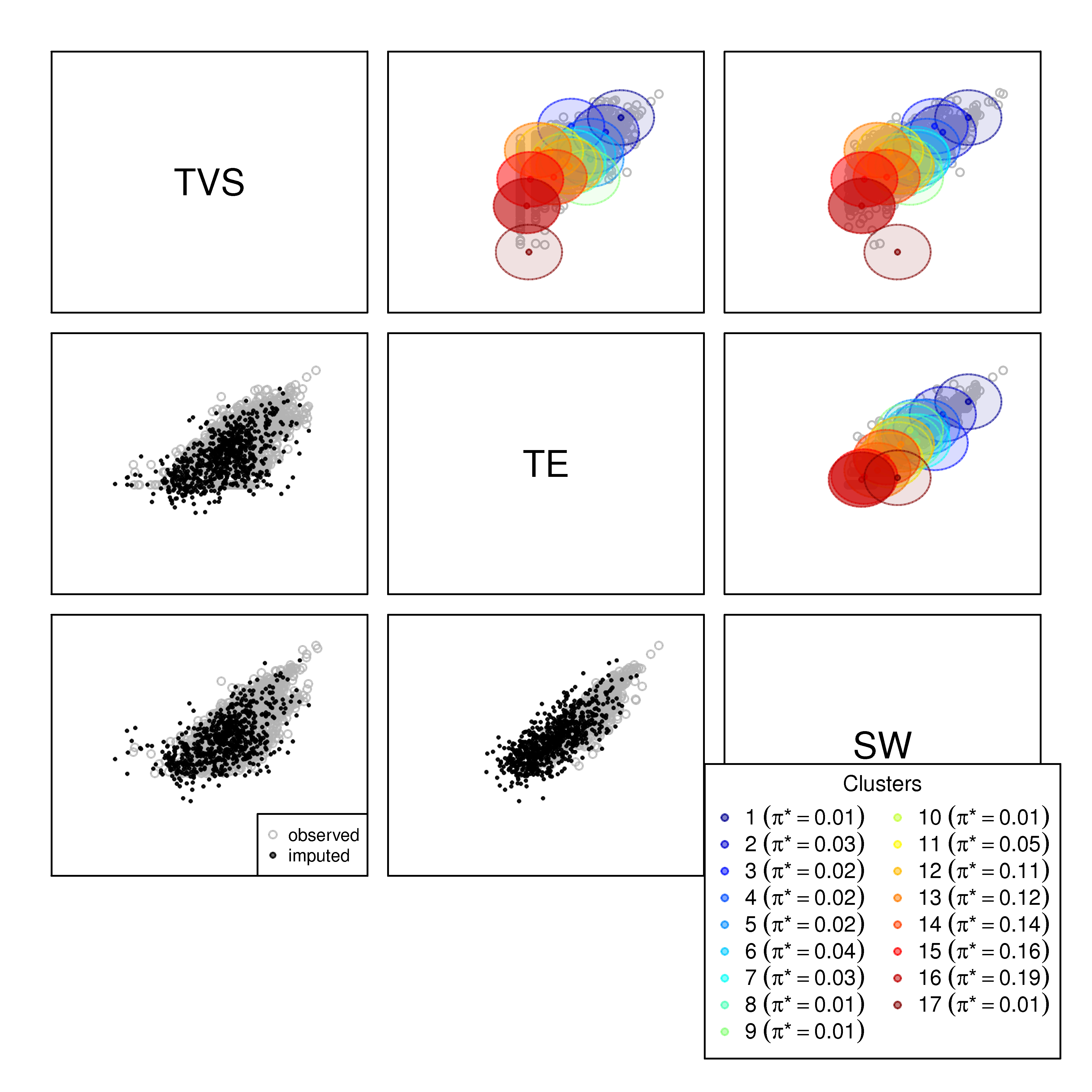}
 \caption[Pairwise scatter plots with the results for the Concrete industry from the 2007 CMF, for the NMAR imputation scenario with higher probabilities for bottom ranked clusters.]{\label{fig:co_bot} Pairwise scatter plots for the concrete industry from the 2007 CMF for the NMAR imputation scenario with higher probabilities for bottom ranked clusters. Observed points are plotted as gray hollow circles. The black filled circles on the lower diagonal are the imputed points. The colored circles on the upper diagonal are the 95\% quantile ellipses of the fitted clusters, with color intensity proportional to the mixture probabilities. Graph copied from the output of the NIMC R tool. Axes removed to prevent disclosure of information about the magnitude of the data.}
\end{figure}

\begin{figure}[t]
 \centering
 \includegraphics[width=\textwidth]{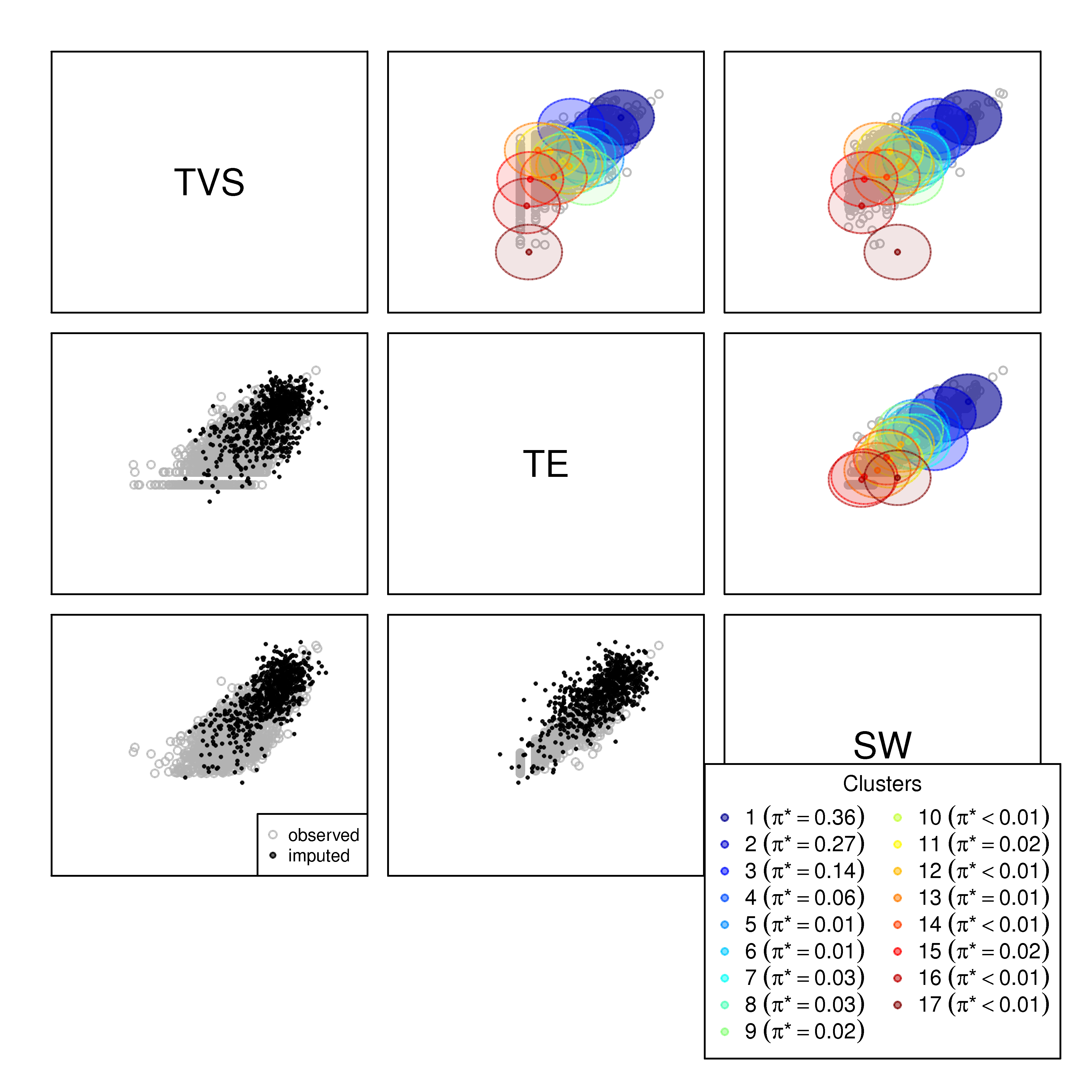}
 \caption[Pairwise scatter plots with the results for the Concrete industry from the 2007 CMF, for the NMAR imputation scenario with higher probabilities for top ranked clusters.]{\label{fig:co_top} Pairwise scatter plots for the concrete industry from the 2007 CMF for the NMAR imputation scenario with higher probabilities for top ranked clusters. Observed points are plotted as gray hollow circles. The black filled circles on the lower diagonal are the imputed points. The colored circles on the upper diagonal are the 95\% quantile ellipses of the fitted clusters, with color intensity proportional to the mixture probabilities. Graph copied from the output of the NIMC R tool. Axes removed to prevent disclosure of information about the magnitude of the data.}
\end{figure}

Figure \ref{fig:co_bot} and Figure \ref{fig:co_top} display results for the concrete industry after imputation for the two NMAR scenarios. In Figure \ref{fig:co_bot}, the imputed values for nonrespondents are primarily in regions with small values of $\bm Y_t$. We accomplish this by setting $\pi_k^m$ to be large for $k \in (11, \dots, 16)$ (the low-ranking components), taking mass away from components in the middle and upper tails of the distribution, e.g., $k \in (4,6)$. We note that the NMAR mechanism still generates large values in the imputed $\mathbf{y}_i$, just not as frequently as would be expected under MAR. This is because we allow $\pi^*_k>0$ for all occupied components, as well as the random chance of imputing values in regions of low density within low-ranking components.

Similarly, in Figure \ref{fig:co_top} we set $\pi_k^*$ to be large for the top-ranking components, generating imputed values for nonrespondents that tend to be in the region of high values of $\bm Y_t$. We defer analysis of the impact of these assumptions on inferences until we present the approach for evaluating stopping decisions.

%------------------------------------------------------------------------------

\section{Using the NMAR Imputations in Stopping Rules} \label{sec:met_adesign}

At any time $t$, the agency needs to decide how many, if any, of the $n_{0} = n - n_t$ nonrespondents to target for follow-up. In this section, we describe how the NMAR imputation approach can be used to inform such decisions. The basic idea is to create completed datasets under various NMAR scenarios, take proposed nonresponse follow-up samples from those completed datasets, and evaluate the cost and accuracy associated with different sample sizes.  This is complementary to the subsampling approach of \citet{Kap14}, with the main difference that we evaluate costs and accuracy with respect to various posited NMAR nonresponse mechanism.

\subsection{Creating completed datasets}

Let $\bm Y_o$ be the data for the $n_0$ unit nonrespondents at time $t$. As these data are unknown, we begin by creating completed versions of $\bm Y = (\bm Y_t, \bm Y_0)$. These serve as plausible ``true values'' that facilitate evaluation of agency nonresponse follow-up decisions. We follow the approach of Section \ref{sec:met_imputation} to create the completed versions, in which we (i) fit the multivariate mixture model using $\bm Y_t$ and (ii) generate several, say $S$, imputation scenarios by adjusting $\bm \pi^*$. Let $\bm \pi^{*s}$ be the adjusted $\bm \pi$ for missingness scenario $s$, where  $s=1, \dots, S$. For each  $\bm \pi^{*s}$, we generate $M$ hypothetical true datasets, $\bm Y^s = (\bm Y^{s,1},\dots, \bm Y^{s,M})$, where $\bm Y^{s,j} = (\bm Y_t, \bm Y_0^{s,j})$ and $\bm Y_0^{s,j}$ for $j=1, \dots, M$ is an independent draw from \eqref{eq:imp_model_zi} and \eqref{eq:imp_model_yi} with $\bm \pi^* = \bm \pi^{*s}$. The use of $S>1$ scenarios allows the agency to assess the implications of nonresponse follow-up decisions under different plausible mechanisms for unit nonresponse. The use of $M>1$ completed datasets allows the agency to propagate uncertainty due to the unit nonresponse and imputation procedure. For simplicity, we impute any missing values in $\bm Y_t$ due to item nonresponse using $\bm \pi^m$ and the parameters from the MAP iteration. 

Let $n_{max} \leq n_{0}$ be the maximum number of nonrespondents that hypothetically could be collected given the total budget, time available, and accuracy requirements for nonresponse follow-up. Here, we are agnostic about how the agency determines $n_{max}$; however, we assume that it is prohibitively expensive to sample all nonrespondents, and the agency prefers to sample fewer cases in follow-up when it can save money without too much sacrifice in accuracy.  
Let $n_{f} = \delta n_{max}$, where $\delta \in[0,1]$, be a number of nonrespondents that the agency is considering to include in a follow-up sample at time $t$. Here, $n_f = 0$ means that the agency stops data collection at time $t$. The agency takes an independent random sample of $n_f$ (possibly zero) individuals from each $\bm Y^{s,j}$. These represent hypothetical nonresponse follow-up samples. After follow-up, the data observed by the agency would include $(\bm Y_t, \bm Y_{f}^{s,j})$, where $\bm Y_{f}^{s,j}$ is the sampled nonrespondents' values. For now, we assume no additional nonresponse in the follow-up sample; we discuss relaxations of this in Section \ref{sec:conclusions}. 

\begin{figure}[t]
\begin{center}

\begin{subfigure}[b]{0.49\textwidth}
 \centering
 \begin{tikzpicture}[inner sep=0, node distance=3cm, auto, bend right=60]
  \tikzstyle{dataset}=[rectangle,draw,minimum height=2cm, minimum width=1.5cm]
  \tikzstyle{post}=[->,shorten >=1pt,semithick]

  \node[dataset, fill=gray!60] (DR) at (0,0) {$\bm{Y}_{t}$};
  \node[dataset, minimum height=1.5cm, fill=gray!30] (DFUS) at (0,-1.75) {$\bm{Y}_f^{s,j}$};
  \node[dataset, minimum height=2.5cm] (DNR) at (0,-3.75) {$\bm{Y}_{0f}^{s,j}$};

  \draw[snake=brace, raise snake=1pt]  (1,1) -- (1,-1) ;
  \draw[snake=brace, raise snake=1pt]  (1,-1) -- (1,-2.5) ;
  \draw[snake=brace, raise snake=1pt]  (1,-2.5) -- (1,-5) ;

  \draw [->] (DFUS.west) to node [left, xshift=-0.2cm] {Imputation} (DNR.west);

  \node [right, xshift=0.3cm](n_r) at (1,0) {$n_t$};
  \node [right, xshift=0.3cm](n_fus) at (1,-1.75) {$n_{f}$};
  \node [right, xshift=0.3cm](n_nr) at (1,-3.75) {$n_{0f}$};

  \node [right, xshift=0.3cm] (N) at (1,-5) {\footnotesize$n = n_t + n_{f} + n_{0f}$};

 \end{tikzpicture} 
 \caption{\label{fig:diag_fusB}}
\end{subfigure}
\begin{subfigure}[b]{0.49\textwidth}
 \centering
 \begin{tikzpicture}[inner sep=0, node distance=3cm, auto, bend right=60]
  \tikzstyle{dataset}=[rectangle,draw,minimum height=2cm, minimum width=1.5cm]
  \tikzstyle{post}=[->,shorten >=1pt,semithick]

  \node[dataset, fill=gray!60] (DR) at (0,0) {$\bm{Y}_{t}$};
  \node[dataset, minimum height=1.5cm, fill=gray!30] (DFUS) at (0,-1.75) {$\bm{Y}_f^{s,j}$};
  \node[dataset, minimum height=2.5cm] (DNR) at (0,-3.75) {$\bm{ Y}_{0f}^{s,j}$};

  \draw[snake=brace, raise snake=1pt]  (1,1) -- (1,-1) ;
  \draw[snake=brace, raise snake=1pt]  (1,-1) -- (1,-2.5) ;
  \draw[snake=brace, raise snake=1pt]  (1,-2.5) -- (1,-5) ;

  \draw[out=-20,in=-160] (DR.north west)  to node (imp) {} (DFUS.south west);
  \draw [->] (imp.west) to node [left, xshift=-0.2cm] {Imputation} (DNR.west);

  \node [right, xshift=0.3cm](n_r) at (1,0) {$n_t$};
  \node [right, xshift=0.3cm](n_fus) at (1,-1.75) {$n_{f}$};
  \node [right, xshift=0.3cm](n_nr) at (1,-3.75) {$n_{0f}$};

  \node [right, xshift=0.3cm] (N) at (1,-5) {\footnotesize$n = n_t + n_{f} + n_{0f}$};

 \end{tikzpicture} 
 \caption{\label{fig:diag_fusA}}
\end{subfigure}
\end{center}  
\caption{\label{fig:diagram_fus} Two possible approaches to estimating the imputation model after hypothetical nonresponse follow-up sampling. The left panel uses only the latest wave of nonrespondents, and the right panel uses all collected data up to time $t$.}
\end{figure}
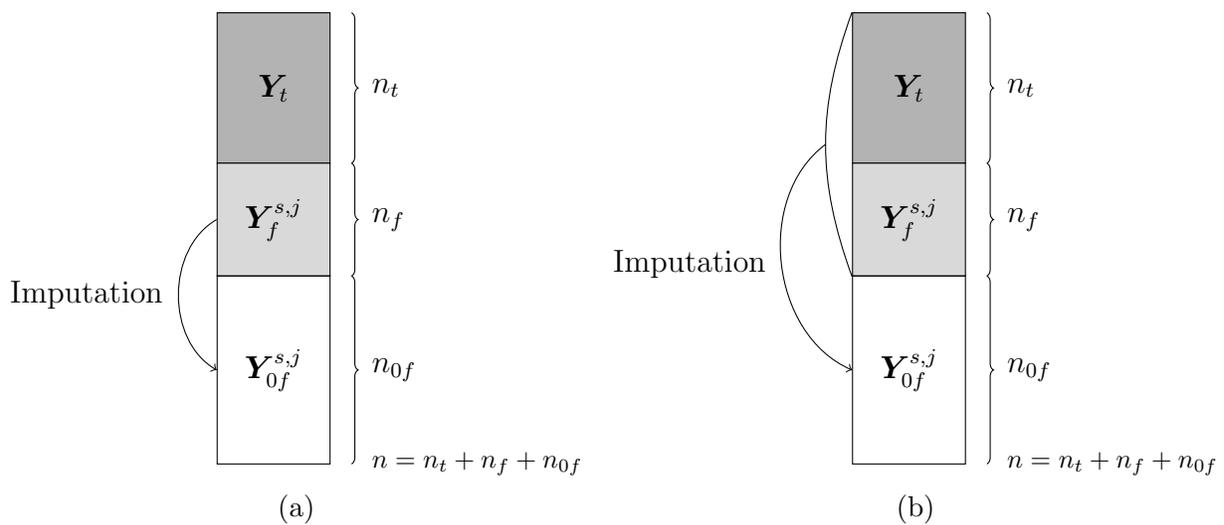

Let $\bm Y_{0f}^{s,j}$ be the data for the $n_{0f} = n_0 - (n_t + n_f)$ individuals that still would be unobserved by the agency after the follow-up sampling. As $\bm Y_{0f}^{s,j}$ would be unknown, following the multiple imputation framework we need to impute plausible values for these $n_{0f}$ cases, thereby creating completed versions of $\bm Y$. To do so, we assume the remaining nonresponse is MAR (even if in reality it is not), which we believe mimics what most agencies would do when faced with imputing nonrespondents' values after completing all follow-up activities. We fit the mixture model again using only the observations in $\bm Y^{s,j}_0$, as depicted in Figure \ref{fig:diag_fusB}, and create $R$ completed datasets by drawing new versions of $\bm Y_{0f}^{s,j}$ from the corresponding posterior predictive distribution. Estimating the imputation model with only $\bm Y^{s,j}_0$ corresponds to the belief that the nonsampled nonrespondents are more similar to the individuals in the latest wave of nonresponse follow-up than to the individuals in $\bm Y_t$. This approach also can be advisable when the follow-up sample is collected under a different protocol than the original sample. When $n_f$ is too small to support reliable modeling, agencies may choose to fit the mixture model to $(\bm Y_t, \bm Y_f^{s,j})$, as shown in Figure \ref{fig:diag_fusA}.

In sum, for each value of $\bm \pi^{*s}$ where $s=1, \dots, S$, we use the following procedure to generate completed versions of $\bm Y$.
\begin{enumerate}
 \item Generate $M$ datasets $(\bm Y^{(s,1)}, \dots, \bm Y^{(s,M)})$ using independent draws of the nonrespondents' values based on $\bm \pi^{*s}$ and the mixture model estimated on $\bm Y_t$. These represent hypothetical ``true values'' of $\bm Y$.
 \item For the value of $\delta$ under consideration and $j=1, \dots, M$, repeat the following steps.
 \begin{enumerate}
  \item Take a random sample of $n_f$ individuals from each $\bm Y_{0}^{(s,j)}$, i.e., from the nonrespondents' ``true values.'' Let $\bm Y_{f}^{(s,j)}$ denote this follow-up sample set.
  \item Fit a new mixture model to $\bm Y_{f}^{(s,j)}$ provided $n_f$ is large enough to support reliable modeling. If not, fit a new mixture model to $(\bm Y_t, \bm Y_{f}^{(s,j)})$.   
  \item Generate $R$ multiply imputed values for the $n_{0f}$ observations that remain nonrespondents, using the posterior predictive distribution of the new mixture model and assuming MAR. For $l=1, \dots, R$, let $\bm Y_{0f}^{(s,j,l)}$ be the $l$th draw of the remaining nonrespondents' values, and let $\bm Y^{(s,j,l)}_{\delta} = (\bm Y_t,  \bm Y_{f}^{(s,j)}, \bm Y_{0f}^{(s,j,l)})$ be a completed version of $\bm Y$ for the specified $\delta$. 
 \end{enumerate}
\end{enumerate}

The agency can repeat Step 2 for multiple values of $\delta$, each time using the same $\{\bm Y^{(s,j)}: s = 1, \dots, S; j = 1, \dots, m\}$. The agency then runs key analyses on $\{\bm Y^{(s,j)}: s = 1, \dots, S; j = 1, \dots, M\}$ and on $\{\bm Y^{(s,j,l)}: s = 1, \dots, S; j = 1, \dots, M; l=1, \dots, R\}$ and evaluates the differences among them, as
we describe next.

\subsection{Utility measures} \label{sec:met_utility}

The decision of stopping the data collection depends critically on the amount of information that has been collected so far, and the amount of information that has yet to be collected. For example, when the unit nonresponse at some time $t$ is close to MAR and $n_t$ is already large, collecting more data may not improve accuracy substantially. On the other hand, when the unit nonresponse is NMAR and $n - n_t$ is large, not sampling the nonrespondents could sacrifice accuracy too much. We propose three measures for assessing accuracy of follow-up sampling choices for multivariate continuous data. The measures do not consider survey weights nor other features of complex designs; we discuss extensions for complex designs in Section \ref{sec:conclusions}.

\subsubsection{Relative percent differences}

For $s=1, \dots, S$,  $j=1, \dots, M$, $v=1,\dots,p$ and $l=1,\dots,R$, let $\bar{y}_v^{(s,j)} = \sum_{i=1}^n y_{i,v}^{(s,j)}/n$ be the marginal mean of the variable indexed by $v$ computed with $\bm Y^{(s,j)}$. Similarly, let $\bar{y}_{v,\delta}^{(s,j,l)} = \sum_{i=1}^n y_{i,v}^{(s,j,l)}/n$ denote the marginal mean of the variable indexed by $v$ computed with $\bm Y_{\delta}^{(s,j,l)}$. For any $s$ and $v$, we compute the average percentage difference, 
\begin{equation} \label{eq:theta}
 \theta_{v,\delta}^{(s)} = \frac{1}{MR} \sum_{j=1}^M  \sum_{l=1}^R
 \left| \frac{\bar{y}_v^{(s,j)} - \bar{y}_{v, \delta}^{(s,j,l)}}{\bar{y}_v^{(s,j)}} \right|.
\end{equation}
Large values of $\theta_{v,\delta}^{(s)}$ are less desirable than small values. When values of $\theta_{v,\delta}^{(s)}$ do not change markedly as $n_f$ increases, higher intensity nonresponse follow-up sampling does not return significant gains in accuracy for variable $v$. We note that one could replace means with any quantity of substantive relevance. As a one number summary, we compute $\theta_\delta^{(s)} = \sum_{v=1}^p \theta_{v, \delta}^{(s)}/p$.

\subsubsection{Difference as fraction of standard error}

For $s=1, \dots, S$,  $j=1, \dots, M$, $v=1,\dots,p$ and $l=1,\dots,R$, let $\hat{\sigma}_v^{(s,j)}$ and $\hat{\sigma}_{v,\delta}^{(s,j,l)}$ denote the empirical standard deviations for variable $v$ computed with $\bm Y^{(s,j)}$ and $\bm Y_{\delta}^{(s,j,l)}$, respectively. To compute each $\hat{\sigma}_v$ we use the square root of the usual formula, $\sum_{i=1}^n (y_{i,v} - \bar{y}_v)^2/(n-1)$. For any $(s, j, l)$ and $v$, we compute the average difference in the means as a fraction of an approximate standard error, 
\begin{equation}
 \tau_{v,\delta}^{(s,j,l)} = \dfrac{\bar{y}_v^{(s,j)} - \bar{y}_{v,
       \delta}^{(s,j,l)}}{\sqrt{\frac{\left[\left(\hat{\sigma}_v^{(s,j)}\right)^2
         + \left(\hat{\sigma}_{v, \delta}^{(s,j,l)}\right)^2\right]\displaystyle/2}{\displaystyle n}}}.
\end{equation}
For any $s$ and $v$, we compute 
\begin{equation} \label{eq:tau}
\tau_{v, \delta}^{(s)} = \frac{1}{MR} \sum_{j=1}^M  \sum_{l=1}^R \left|\displaystyle \tau_{v,\delta}^{(s,j,l)}\right|,
\end{equation}
Like $\theta_{v,\delta}^{(s)}$, this criterion penalizes large differences in the means of $\bm Y^{s,j}$ and the completed datasets. However, it considers differences in terms of standardized units. As a one number summary, we compute
$\tau_\delta^{(s)} = \sum_{v=1}^p \tau_{v, \delta}^{(s)}/p$.

\subsubsection{Propensity score distances}

\citet{Woo09} present a metric for quantifying the similarity of the multivariate empirical distributions in two datasets. They develop this metric to evaluate the analytic validity of redacted data, comparing the empirical distributions in the redacted and corresponding confidential data. We adapt their metric for nonresponse follow-up decisions as follows. For each $(s, j, l)$, we concatenate $\bm Y^{(s,j)}$ and $\bm Y_{\delta}^{(s,j,l)}$. We add a variable $T_{i}=1$ for all rows belonging to $\bm Y^{(s,j)}$ and $T_{i}=0$ for all rows belonging to $\bm Y_{\delta}^{(s,j,l)}$. We then regress $T$ on main effects for $(y_1, \dots, y_p)$ using a a generalized additive model estimated on the concatenated data \citep{Woo08}. With this regression, we compute the predicted probability for each observation, $\hat{e}_{i, \delta}^{(s,j,l)}$. For any $(s,j,l)$, we compute 
\begin{equation} \label{eq:rho}
 \rho_{\delta}^{(s,j,l)} = \frac{\sum_{i=1}^{2n}(\hat{e}_{i, \delta}^{(s,j,l)} - 0.5)^2}{2n}. 
\end{equation}
This formula is related to the R indicator \citep{Sch09}.
When $\bm Y^{(s,j)}$ and $\bm Y_{\delta}^{(s,j,l)}$ have similar empirical distributions, the values of $\hat{e}_{i, \delta}^{(s,j,l)}$ should be close to 0.5, so that $\rho_{\delta}^{(s,j,l)}$ should be near zero \citep{Woo09}.

As a metric, we compute the average,
\begin{equation} \label{eq:rho}
 \rho_{\delta}^{(s)} = \frac{1}{MR} \sum_{j=1}^M  \sum_{l=1}^R \rho_{\delta}^{(s,j,l)}. 
\end{equation}
When $\rho_{\delta}^{(s)}$ is close to zero, the ``true value'' and multiply-imputed databases contain similar information content; in this case, nonresponse follow-up may not be worth the cost. When $\rho$
is close to $1/4$, the ``true value'' and multiply-imputed databases have quite different distributions, suggesting that there is information to be gained by additional follow-up sampling. \citet{Woo09} do not suggest values of $\rho$ that can be considered ``good enough.'' Thus, we recommend comparing relative values of $\rho$ for different follow-up sample sizes and scenarios.

\subsection{Cost measure} \label{sec:met_cost}

As $n_f$ increases, $\bm Y^{(s,j)}$ and corresponding $\bm Y_{\delta}^{(s,j,l)}$ have increasing numbers of overlapping records, causing the measures of Section \ref{sec:met_utility} to decrease. Of course, agencies must balance accuracy gains with increased expenses. Therefore, we consider introducing cost functions into the decision process, so that agencies can decide how much they are willing to spend for utility improvements. Here, we adopt a simple cost function for illustration, primarily to emphasize the importance of considering cost in the decisions. For any $(s,j)$, let $C_f^{(s,j)}$ be the total cost of the follow-up sample, $C_0^{(s,j)}$ be the fixed cost regardless of the sample size, and $c_i^{(s,j)}$ the cost of selecting, measuring and processing unit $i$. Many types of fixed costs can be folded into the fixed costs of the original survey; these can be excluded from $C_0^{(s,j)}$. We keep $C_0^{(s,j)}$ in the formulation in case there are fixed costs associated with follow-up; however, it easily could be set to zero. We assume the cost function,
\begin{equation} \label{eq:cost}
C_f^{(s,j)} = C_0^{(s,j)} + \sum_{i \in \bm Y_f^{s,j}} c_i^{(s,j)}.
\end{equation}
For the illustration with CMF data, we assume that $c_i^{(s,j)}=c$ for all nonrespondents and all $(s,j)$; that is, we make nonresponse follow-up costs a linear function of $n_f$. We also assume that $C_0^{(s,j)} = C_0$ for all $(s,j)$. Under these assumptions, $n_{max}$ is the largest value of $n_f$ such that $C_0 + c n_f$ is less than the total budget available for nonresponse follow-up at time $t$.

The framework readily handles more complex cost functions \citep[for examples, see][]{Gro04}. When agencies allow $c_i^{(s,j)}$ to vary across individuals or scenarios, one can add in \eqref{eq:cost} the realized cost for each individual sampled in Step 2(a) of the algorithm from Section \ref{sec:met_adesign}.

%------------------------------------------------------------------------------

\section[Illustration with Census of Manufactures Data]{Nonresponse follow-up Decisions in the Census of Manufactures Data} \label{sec:application}

The U.\ S.\ Census Bureau spends significant resources on nonresponse follow-up operations during the CMF data collection. For example, they mail additional questionnaires to nonrespondents and designate Census consultants to work with some of the largest establishments. The entire data collection process takes more than a year to complete. Thus, the Census Bureau potentially could benefit from sensible stopping decisions that reduce costs and shorten time to data release without too much sacrifice in accuracy. 

To illustrate the methodology, we use the data and missingness scenarios described in Section \ref{sec:imp_census}. For brevity, we report only results for the concrete industry. The results for the plastics products industry can be found in the supplementary material. We consider $\delta \in (0, 0.25, 0.5, 0.75, 1)$, that is, collecting follow-up samples with 0\%, 25\%, 50\%, 75\% and 100\% of $n_{max}$. We set $n_{max}=n_{0}$, corresponding to sufficient resources to attempt to collect all nonrespondents' data if desired. For each scenario $s$, we create $M=10$ completed hypothetical censuses in the two industries. For each $\bm Y^{(s,j)}$ and for each value of $\delta$, we take simple random follow-up samples to obtain hypothetical versions of $\bm Y_{f,\delta}^{(s,j)}$. We generate $R=5$ completed datasets $\bm Y_{\delta}^{(s,j,l)}$ by fitting new mixture models to $\bm Y_{f,\delta}^{(s,j)}$ only; for comparison, we also report results with mixture models based on $\left(\bm Y_t, \bm Y_{f,\delta}^{(s,j)}\right)$. Due to disclosure limitation requirements, at the request of the Census Bureau we do not report values for the cost measure. Since the cost measure is a linear function of $n_f$, we believe that we can illustrate the decision-making process and sensitivity analysis using results based on $\delta$.    

\begin{figure}[t]
 \centering
 \includegraphics[height=0.3\textheight]{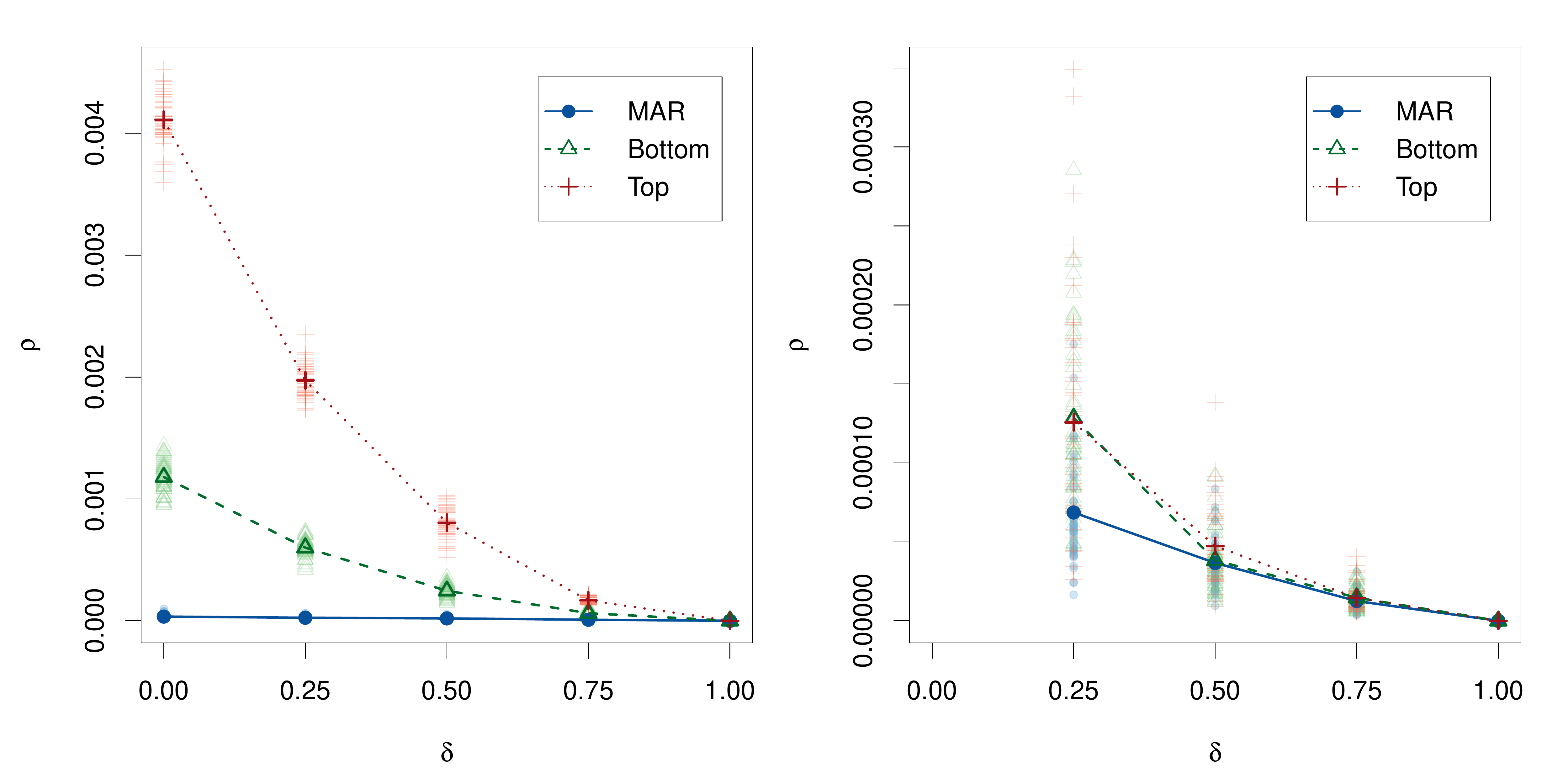}
 \includegraphics[height=0.3\textheight]{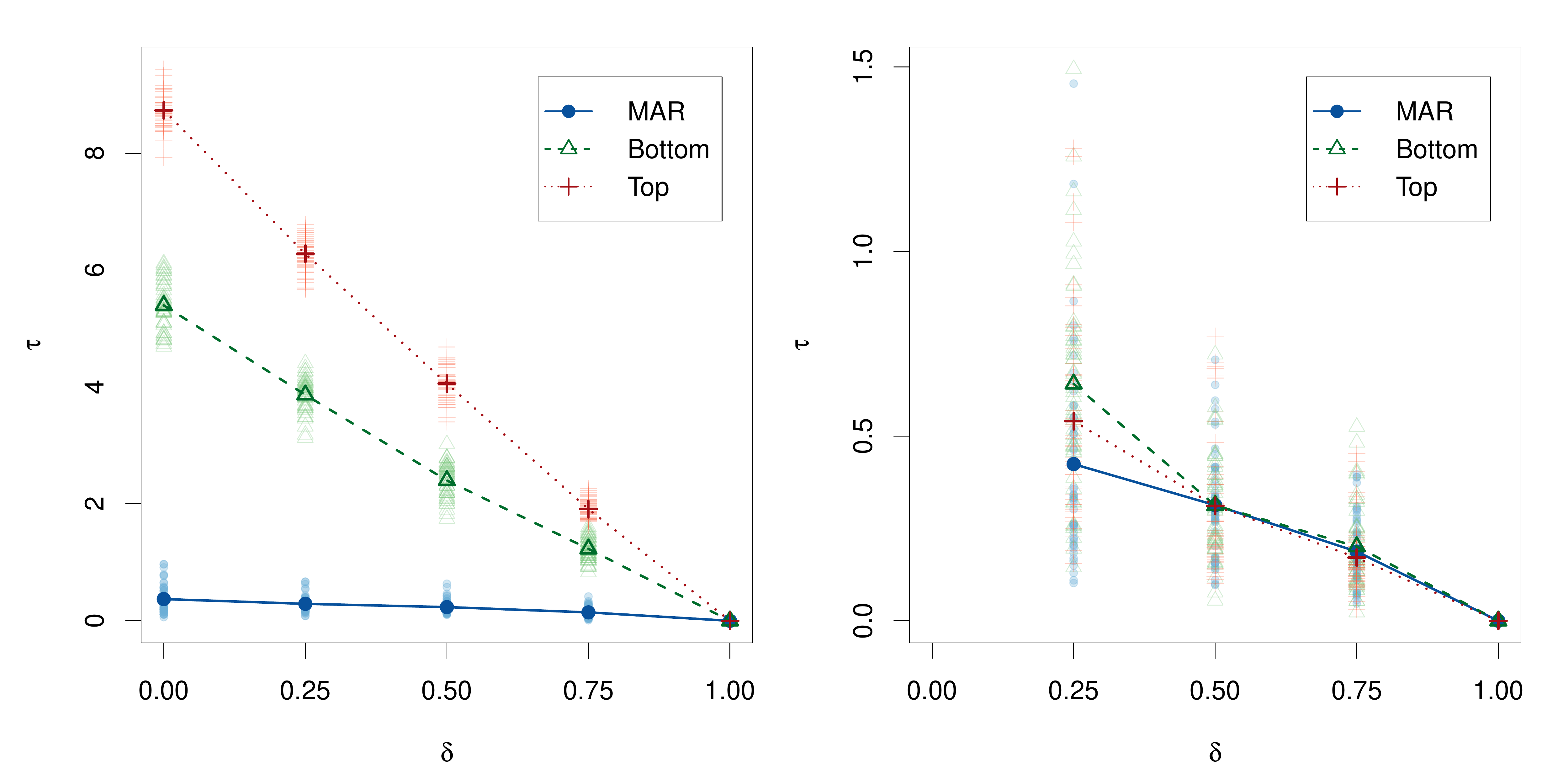}
 \includegraphics[height=0.3\textheight]{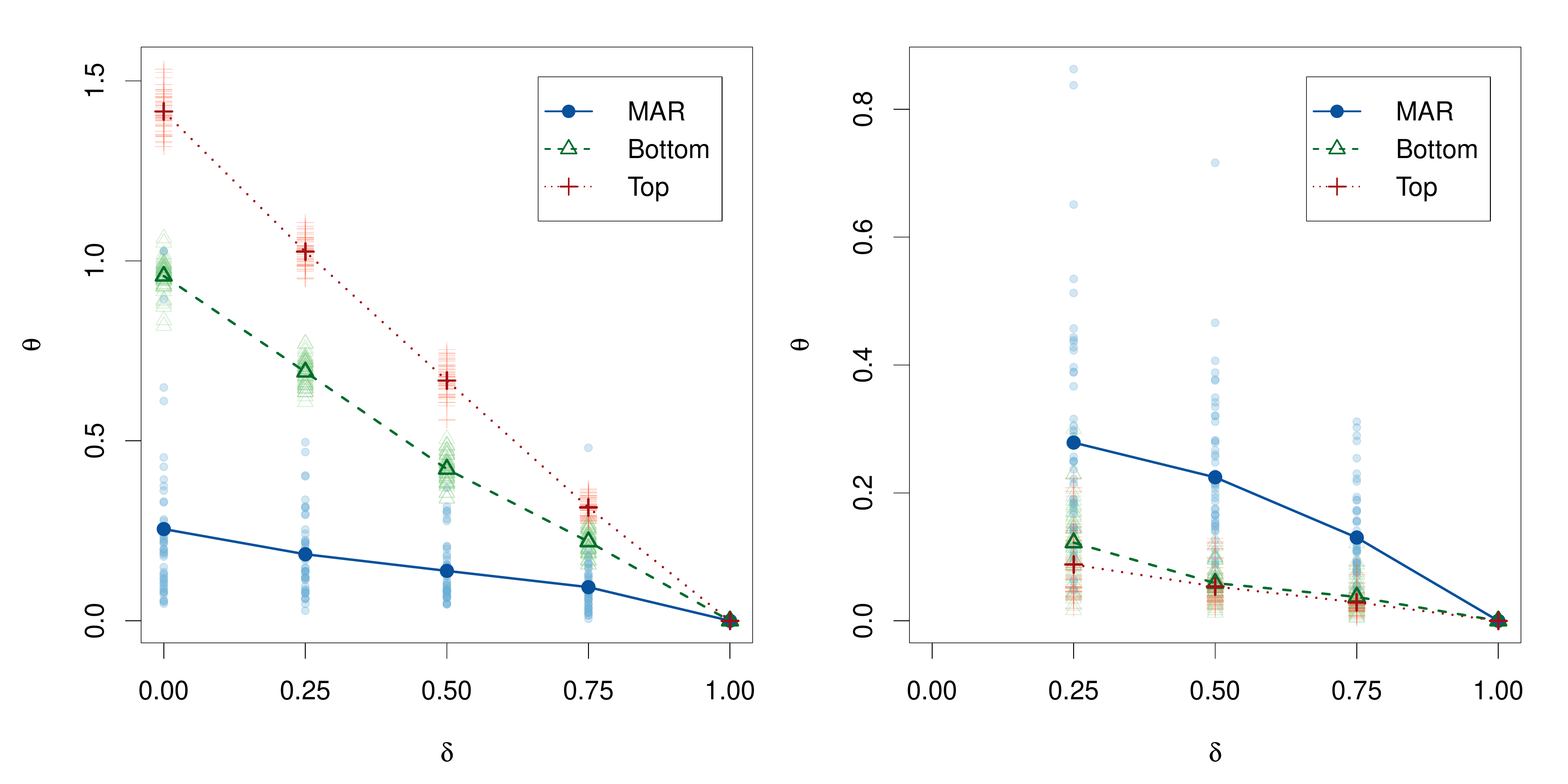}
 \caption[Summary of utility measures for the three scenarios considered for the Concrete industry from the 2007 CMF.]{\label{fig:measures_concrete}Summary of utility measures for the three scenarios for the concrete industry from the 2007 CMF. The results for $\rho$, $\tau$ and $\theta$ are on the first, second and last row, respectively. The plots on the left column display the results with the model fit to observed data and follow-up sample, and the plots on the right display the results with the model fit to follow-up sample only. The faded points are the individual values for each multiple imputation. 
}
\end{figure}

Figure \ref{fig:measures_concrete} displays $(\theta_\delta^{(s)}, \tau_\delta^{(s)}, \rho_\delta^{(s)})$ for the three missingness scenarios in the concrete industry data as a function of $\delta$. As expected, in all cases the measures decrease as $\delta$ increases. When $\delta=1$, all units are sampled so that all utility measures equal zero. When $\delta=0$, imputations are only based on $\bm Y_t$. In the MAR scenario, we produce greater accuracy when generating the $R=5$ imputed datasets from the model estimated with $\left(\bm Y_t, \bm Y_{f,\delta}^{(s,j)}\right)$ than the model estimated with $\bm Y_t$. This is expected, since the former model is based on more observations than the latter, and both $\bm Y_t$ and $\bm Y_{f,\delta}^{(s,j)}$ are representative of the nonrespondents under MAR. In contrast and also as expected, in the NMAR scenarios the values of the utility measures are much smaller when the imputation models are estimated only with $\bm Y_{f,\delta}^{(s,j)}$, which is representative of the nonrespondents whereas $\bm Y_t$ is not. The values of the utility measures are generally smaller for the MAR scenario than for the NMAR scenario. The values in the MAR scenario can serve as baselines, providing a sense of the magnitudes of ``best case'' results.  

Figure \ref{fig:measures_concrete} can be used to make qualitative decisions about nonresponse follow-up. When the agency suspects the nonrespondents' data are close to MAR, conducting a follow-up sample does not improve accuracy dramatically with respect to these measures. For example, a 25\% nonresponse follow-up sample decreases $\theta_\delta^{(s)}$ by around 7 points (from .255 to .185), and a 50\% nonresponse follow-up sample decreases $\theta_\delta^{(s)}$ by around 11.6 points (from .255 to .139). In contrast, when the agency is concerned that the data are NMAR in ways defined by the two selected $\bm \pi^{*s}$ (with the vectors of values shown in the legend of Figures \ref{fig:co_bot} and \ref{fig:co_top}), collecting data from 25\% or 50\% of the nonrespondents results in massive accuracy improvements: $\theta_\delta^{(s)}$ drops from values of at least 1.0 when $n_f = 0$ to values in line with those for the MAR scenario. For all scenarios, beyond $\delta = .50$ the gains seem unlikely to be worth the cost. We reach similar qualitative conclusions for all three utility measures. 

Agencies may want to plug utility and cost measures into specific loss functions, and find $n_f$ that minimize loss. We leave specification of suitable loss functions to future work.

%--------------------------------------------------------------------------
% Figures Concrete 

\clearpage

%-----------------------------------------------------------------------------

\section{Concluding Remarks} \label{sec:conclusions}

It is prudent for agencies to consider several plausible mechanisms for unit nonresponse when assessing sensitivity of inferences to NMAR nonresponse and making decisions about follow-up. In some settings, it may be possible to use previous data collections to inform those mechanisms. For example, the CMF is collected every five years. It may be reasonable to assume that differences between respondents and nonrespondents at similar time points $t$ are similar across surveys. When this is the case, the agency might mimic patterns seen in the previous data collection to make scenarios in the current data collection. For example, suppose the agency estimates the mixture model using data from the earlier data collection up to time $t$, and finds that respondents after time $t$ were most likely to come from a certain region of the variable space. For the current data collection, the agency could increase the mixture probabilities for the components in that space.

With multiple mechanisms, agencies can ease interpretations by using expected values of the utility and cost measures. To compute these, agencies can put a subjective probability $p_s$ on each missing data scenario $s$ under consideration, representing the agency's belief on the chance of each scenario being (close to) reality. The agency can average the utility measures and costs over the subjective probabilities, e.g., $\sum_s \theta_{\delta}^{(s)} p_s$. 

In contrast, evaluations based on highly unlikely NMAR scenarios can be misleading. For example, suppose that the nonresponse mechanism is truly MAR, but an agency puts heavy weight (unwisely) on a scenario with very different distributions for nonrespondents and respondents. In this case, the decision is likely to lead to unnecessarily large follow-up samples, in that a MAR model applied to the data at hand might be sufficiently accurate for a relatively low cost. On the other hand, when the true nonresponse mechanism is NMAR but the agency assumes otherwise, the evaluations may suggest not to take a large follow-up sample, which could leave the agency with nonrepresentative data, even after imputation. This emphasizes the importance of feeding only plausible scenarios into this, or any, decision making process.

In any follow-up sample, it is likely that some sampled individuals will continue to be nonrespondents. To account for this in the decision-making process, agencies can propose models for the nonresponse in the follow-up samples. For example, the agency can randomly sample nonrespondents within the follow-up sample from a specified selection model. To mimic typical practice and simplify the imputation process, the agency could combine any nonrespondents in the follow-up sample with the nonsampled cases, and use MAR models based on respondents in $\bm Y_{f,\delta}^{(s,j)}$ to create the completed $\bm Y_{\delta}^{(s,j,l)}$. Investigating how such additional modeling affects decisions about $n_f$ is a topic for future research.

We present the approach for deciding nonresponse follow-up sample size for a single time point. However, agencies could adapt the methodology for multiple decision points. For example, in the CMF the Census Bureau collects and processes data in waves during the year, so that it has the opportunity to adapt nonresponse follow-up designs at multiple time points points, e.g., monthly or quarterly. The agency can reset the process at each $t$, using the collected data up to that time as $\bm Y_t$ and recreating hypothetical nonrespondents' data. In this case, agencies may need to re-consider $n_{max}$ at each wave, for example, based on available budget and required accuracy levels.

The methodology is directly applicable to census data, but it could be modified for surveys. They key issue is to model relationships between the survey design variables and the substantive survey variables. We conjecture that one could regress on functions of the design variables when specifying the means of the mixture components. Alternatively, a more {\em ad hoc} means to do so is to include the survey weights as a variable in the mixture model. Developing imputation measures, as well as utility measures and cost functions, that explicitly account for complex design features is a topic for future research.

\section{Acknowledgments} 

%The authors thank Kirk White of the U.\ S.\ Census Bureau for
%assistance with accessing the Census of Manufactures and with
%imputation of the missing data. 
%This research was supported by the NSF NCRN grant (SES-11-31897)
%awarded to Duke University. 
Any opinions and conclusions expressed in this article are those of the authors and do not necessarily represent the views of the U.\ S.\ Census Bureau. All results have been reviewed to ensure that no confidential information is disclosed.

%-----------------------------------------------------------------------------

\bibliographystyle{apalike}
\bibliography{paper_confid,disclosurebib,nonparametrics,adesign,imputation}

%-----------------------------------------------------------------------------

\clearpage

\end{document}